\documentclass{svjour2}                    % onecolumn
\smartqed  % flush right qed marks, e.g. at end of proof
\usepackage{graphicx}
\newcommand{\binom}[2]{\left(\!\! \begin{array}{c}{#1}\\ {#2} \end{array}\!\!\right)}

\begin{document}

\title{
Realization of the open-boundary totally asymmetric simple exclusion process on a ring
}

\author{Masahiro Kanai}

\institute{M. Kanai \at
Graduate School of Mathematical Sciences, The University of Tokyo, Komaba 3-8-1, Meguro-ku, Tokyo, Japan\\
              Tel.: +81-3-5465-8310\\
              \email{kanai@ms.u-tokyo.ac.jp}
}

\date{Received: date / Accepted: date}

\maketitle

\begin{abstract}
We propose a misanthrope process, defined on a ring, which realizes the totally asymmetric simple exclusion process with open boundaries.
In the misanthrope process, particles have no exclusion interactions in contrast to those in the simple exclusion process, while the hop rates depend on both numbers of particles at departure and arrival sites.
Arranging the hop rates, we can recover the simple exclusion property and also make a condensate, which grows at an arbitrary single site, behave as an external reservoir providing and absorbing particles.
It is known that, under some condition, the misanthrope process has an exact solution for its steady-state probability.
We exploit this fact to see in an analytical way that the model proposed is exactly what we expected.
\end{abstract}

%\begin{figure}
%  \includegraphics{example.eps}
%\end{figure}
%\begin{figure*}
%  \includegraphics[width=0.75\textwidth]{example.eps}
%\end{figure*}
%\begin{table}
%\begin{tabular}{lll}
%\hline\noalign{\smallskip}
%\end{tabular}
%\end{table}

%%%%%%%
\section{Introduction}
%%%%%%%
Since proposed independently in 1968 \cite{MGP} and 1970 \cite{Spitzer}, the totally asymmetric simple exclusion process (TASEP) has so far been a cornerstone of interactions among mathematics \cite{KL,Liggett,Spohn,KNT06}, nonequilibrium statistical physics \cite{SZ,Schutz,Derrida07}, biological transport \cite{SEPR,KH,CMZ,ZDS}, traffic flow \cite{CSS,CSS2,Popkov,Kanai05} and so on.
In the early 1990's, two key studies on the TASEP with open boundaries were published:
one reveals for the first time that nonequilibrium phase transitions, induced by open boundaries, occur in the TASEP \cite{Krug}; the other gives a mathematical method to obtain the steady-state probability in an analytical form for the TASEP, which is now referred to as the matrix-product ansatz \cite{DEHP}.
Because of the simple definition of TASEP and the versatility of the matrix method, extensive studies on particle-hopping models have followed thus far (see \cite{BE} and the references therein).
Evans M. R., one of the authors of \cite{DEHP}, has been exploring the area of exactly solvable models such as the zero-range process \cite{Spitzer,EH,Kanai07} and the misanthrope process \cite{Cocozza,EW,Kanai10}.
We note that these two processes are usually defined on a ring.

In this work, we do a trick of realizing open boundaries on a 1-dimensional periodic lattice as follows; the misanthrope process defined on a ring realizes the TASEP with open boundaries.
First, we recall the TASEP with open boundaries: a particle comes into the finite lattice at the leftmost site and then hops to the right neighboring site with a given hop rate before finally getting out of the lattice from the rightmost site.
Also, particles in the lattice obey volume-exclusion interactions, i.e., each site contains at most one particle.
We need to make the misanthrope process incorporate the exclusion property as well as the boundary and hop rates.

The key is that the hop rate of particles in misanthrope process depends not only on the particle number of the departure site but also on that of the arrival site.
Arranging the hop rate in misanthrope process, although the particles do not have any exclusion interaction, we succeed in mimicking the exclusion property and moreover making one of the sites selected to play the role of both boundaries.
More precisely, in the steady state a condensate grows at an arbitrary site, providing a particle to the right-hand site and also absorbing a particle from the left-hand site as if it were to be an external reservoir of particles.
Note that the present model is different from such models as a TASEP appended by a special site subject to quite different rules \cite{ASZ,CZ}.

The advantage of the misanthrope process is that it has an exact steady-state probability when certain conditions on the hop rates are satisfied \cite{Cocozza}.
It is fortunate that we can rearrange the hop rates in the present model so as to satisfy the conditions for the exact steady state.
We can thus confirm that the present model successfully behaves as we expected.

This paper is organized as follows.
In Section 2, we explicitly define the models mentioned above, i.e., the TASEP and the misanthrope process.
Then we describe our model, which is a misanthrope process with the hop rates designed for the present purpose.
In Section 3, we show simulation results on the model proposed, which successfully recover phenomena well-known in the TASEP with open boundaries.
In Section 4, we modify our model so that it allows for exact solutions, and thus show some analytical results to see that the present model is what we expected.
Section 5 is devoted to a summary and conclusions.

%%%%%%%%%%
\section{TASEP and the misanthrope process}
%%%%%%%%%%
We now define the models: TASEP and the misanthrope process.
%Consider a 1-dimensional lattice of $L$ sites on which $N$ particles reside hopping in the same, definite direction (say, to the right).
In both models, we consider a 1-dimensional lattice of $L$ sites on which some particles reside hopping in the same, definite direction (say, to the right).
%Each site $l$ carries $n_l$ particles, so that $\sum^L_{l=1}n_l=N$.
Each site $l$ carries $n_l$ particles; in the TASEP, $n_l$ takes the value of 0 or 1 due to the exclusion interactions of particles, while it may take the value of 0 or positive integer in the misanthrope process.
A particle in the TASEP hops from site $l$ to site $l+1$ with a constant rate $p$ (which may be normalized to be 1), while one does with rate $u(n_l,n_{l+1})$ in the misanthrope process.
To be precise, the hop rates are per unit time $\Delta t$, i.e., particles hop with probability $p\Delta t$ or $u(n_l,n_{l+1})\Delta t$.
Accordingly we consider the random sequential updating scheme in simulation.

%%%%%%
\begin{figure}[htb]
\begin{center}
\includegraphics[scale=0.4]{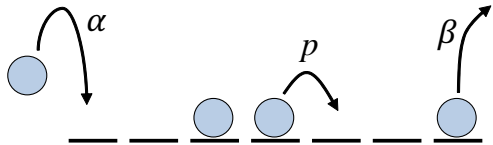}
\caption{
The TASEP with open boundaries.
Particles come into the lattice from outside at the leftmost site with rate $\alpha$, hopping to the right with rate $p$ under the exclusion interaction and getting out of the lattice from the rightmost site with rate $\beta$.
}
\label{TASEP}
\end{center}
\end{figure}
%%%%%%
In the TASEP, we assume the open boundary condition: a particle comes into the lattice with rate $\alpha$ at the leftmost site $1$ if $n_1=0$; one gets out of the lattice with rate $\beta$ as it reaches the rightmost site $L$.
The number of particles in the lattice, as a result, fluctuates at every moment.
In Fig. \ref{TASEP}, we illustrate the TASEP with the open boundary conditions.
In the misanthrope process, we consider a ring geometry, i.e., the periodic boundary condition: a particle hops from site $L$ to site 1 with rate $u(n_L,n_{L+1})$ where $n_{L+1}=n_1$.
Accordingly, there is no special site in the lattice from which a particle hops with a different rate.
In the misanthrope process, we denote by $N$ the total number of particles conserved: $N=\sum^L_{l=1}n_l$.

%%%%%%%%%%%
\begin{table}[htb]
\begin{center}
\begin{tabular}{c|ccccc}
$u(m,n)$ & $n=0$ & $n=1$ & $n=2$ & $n=3$ & $\cdots$ \\
\hline
$m=0$ & $0$ & $0$ & $0$ & $0$ & $\cdots$ \\
$m=1$ & $p$ & $0$ & $\beta$ & $\beta$ & $\cdots$ \\
$m=2$ & $\alpha$ & $0$ & 1 & 1 & $\cdots$ \\
$m=3$ & $\alpha$ & $0$ & 1 & 1 & $\cdots$ \\
$\vdots$ & $\vdots$ & $\vdots$ & $\vdots$ & $\vdots$ & $\ddots$
\end{tabular}
\caption{
Hop rates in the misanthrope process that we consider here.
Note that $u(1,0)=p$ corresponds to the hop rate of particles on the lattice in TASEP.
The second column $u(m,1)=0$ means the exclusion interaction of particles.
$u(m\geq2,0)=\alpha$ and $u(1,n\geq2)=\beta$ correspond to the boundary rates in the TASEP.
%When a site carries more than 1 particle, it provides a particle to the right site with rate $u(m \geq 2, 0)=\alpha$ and absorbs one from the left site with rate $u(1,n\geq2)=\beta$.
We assume that $p$, $\alpha$ and $\beta$ are positive.
}
\label{hoprate}
\end{center}
\end{table}
%%%%%%%%%%%
In Table \ref{hoprate}, we build the hop rates $u(m,n)$ so that the misanthrope process mimics the TASEP with the open boundary condition defined above.
It is natural that one should take all hop rates to be positive or zero.
$u(1,0)=p$ corresponds to the hop rate of particles on the lattice in TASEP.
The exclusion interaction of particles in the TASEP is given by $u(m,1)=0$, which means that no particle can enter sites already occupied by another one.
In contrast, when a site carries more than 1 particle, it starts to provide a particle to the right site, if empty, with rate $u(m \geq 2, 0)=\alpha$ as well as to absorb one from the left site with rate $u(1,n\geq2)=\beta$.
We do not have to determine the other rates exactly but take $u(m\geq2,n\geq2)=1$ for simplicity.
This choice strongly urges the sites to discharge their surplus particles if there are consecutive sites carrying more than 1 particle.

%%%%%%%%
\section{Simulation results}
%%%%%%%%

%%%%%%%%%%%
\begin{figure}[htb]
\begin{center}
\includegraphics[scale=.8]{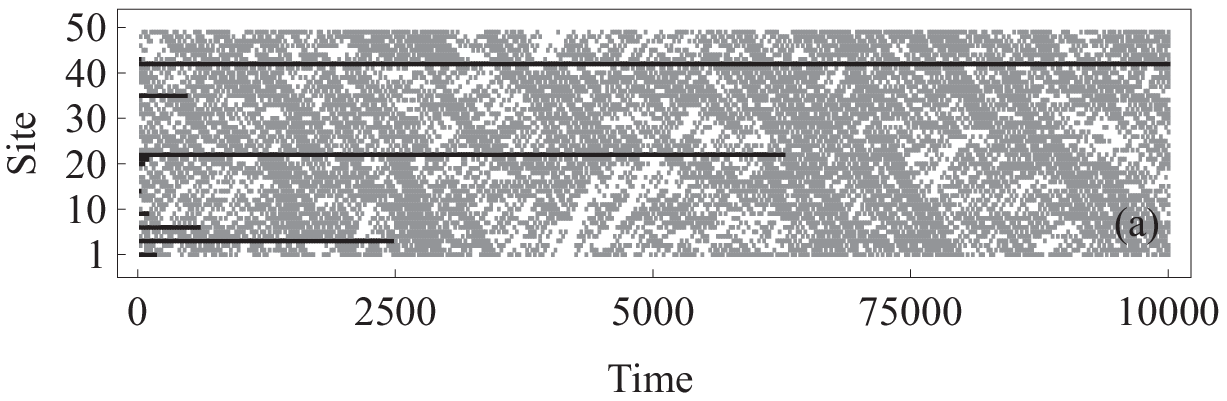}
\includegraphics[scale=.8]{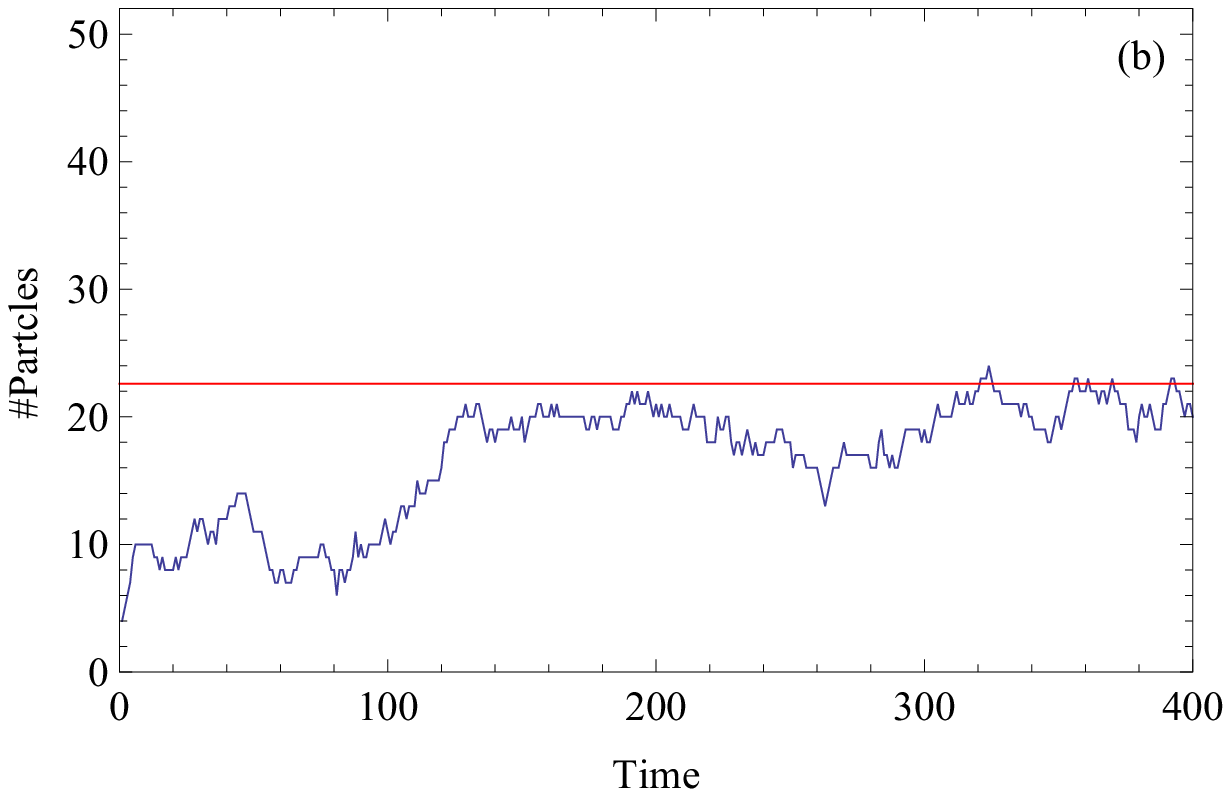}
\includegraphics[scale=.8]{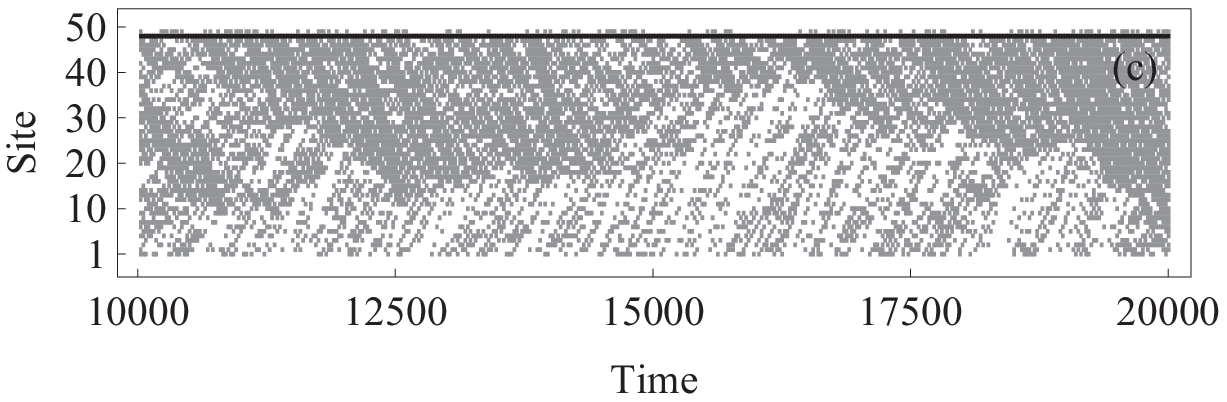}
\caption{Simulation results for the misanthrope process with the hop rates given in Table \ref{hoprate}: $L=50$, $N=52$, $p=1$.
(a)Spatio-temporal pattern: $\alpha=0.6$ and $\beta=0.4$.
The number of particles at each site is displayed; gray sites carry one particle and black sites do more than one.
Particles move from the bottom to the top; the top edge connects with the bottom edge to be periodic.
(b)Maximum number of the particles accumulated at a single site:
it corresponds to the height of the black sites in Fig. (a).
The horizontal line indicates the exact value of $N-(L-1)(1-\beta)=22.6$ (discussed in Section \ref{Analytical}).
(c)Shock profile: $L=50$, $N=52$, $\alpha=\beta=0.3$.
}
\label{fig}
\end{center}
\end{figure}
%%%%%%%%%%%

In this section, we show simulation results for the misanthrope process with the hop rates given in Table \ref{hoprate}.
Figure \ref{fig}(a) shows the spatio-temporal pattern of the model.
Starting from a random configuration, in which particles are randomly distributed across the lattice, there remain three sites each carrying more than 1 particle for a while.
Eventually these sites disappear except one.
Figure \ref{fig}(b) shows the maximum number of particles accumulated at a single site at each time. 
Figure \ref{fig}(c) shows a shock profile of density.
One finds that the shock front randomly fluctuates, separating the lattice into dense and sparse domains.

The phase diagram with respect to the boundary rates $\alpha$ and $\beta$ has some distinct phases: high-density, low-density, maximal-current and shock phases.
The diagram of the TASEP with open boundaries has been exactly obtained \cite{Krug,DEHP,Derrida92}.
In the limit that the number of sites $L$ tends to infinity, the density $\rho$ in the bulk (i.e., at the sites far from both ends of lattice) and the current $Q$ are explicitly given as follows (where $p$ is normalized to be $1$ without loss of generality):
%(i)$\alpha \geq p/2$ and $\beta \geq p/2$: $\rho=1/2$ and $Q=p/4$, (ii)$\alpha < p/2$ and $\beta > \alpha$: $\rho=\alpha/p$ and $Q=\alpha(1-(\alpha/p))$, (iii)$\beta < p/2$ and $\alpha > \beta$: $\rho=\alpha/p$ and $Q=p/4$.
(i)$\alpha \geq 1/2$ and $\beta \geq 1/2$ (maximal-current phase): $\rho=1/2$ and $Q=1/4$, (ii)$\alpha < 1/2$ and $\beta > \alpha$ (low-density phase): $\rho=\alpha$ and $Q=\alpha(1-\alpha)$, (iii)$\beta < 1/2$ and $\beta < \alpha$ (high-density phase): $\rho=1-\beta$ and $Q=\beta(1-\beta)$, (iv)$\alpha=\beta<1/2$ (shock phase): there coexist high-density and low-density profile to arise a shock in the middle of lattice.
The shock phase is a first-order phase transition line separating the low-density and high-density phases.
Actually, the shock front performs a random walk, being reflected from the ends.
See \cite{SD} for more detailed classification of the phase diagram, and \cite{Derrida98,SA} for the shock phase.

Working on simulation, we can see that the present model actually recovers these analytical results in a whole range of parameters as far as the system size ($L$ and $N$) is small; nevertheless, it takes much longer time for the system to reach the steady state as the size increases.
In particular, it may be difficult for one to find how the steady state will be in the case that $N=L$ and both are large.
In the next section, we alternatively investigate the model in an analytical manner under the constraint $\alpha+\beta=p$ that allows us to have an exact steady-state probability;
with the use of the exact solution, we shall calculate the expectation values of density $\rho$ and current $Q$ to compare with the above results.
Note that one can take a sample of $\alpha$ and $\beta$ from each of the phases (i), (ii) and (iii) because the line $\alpha+\beta=p$ gets across them.
Therefore, it is sufficient for the present discussion to see that the misanthrope process succeeds in realizing the TASEP with open boundaries for any $\alpha$ and $\beta$ taken under the above constraint.

%%%%%%%%%%%
%\begin{figure}[htb]
%\begin{center}
%\caption{condensate of 12 particles; $L=10$ $N=18$ $\alpha=0.6$ $\beta=0.4$}
%\label{cond}
%\end{center}
%\end{figure}
%%%%%%%%%%%

%%%%%%%%
\section{Analytical results}\label{Analytical}
%%%%%%%%
We focus on the steady states of the misanthrope process in which the probability of a configuration to occur is independent of time.
The misanthrope process has an exact solution of the steady-state probability.
Due to this remarkable property, we can calculate expectation values for the process in an analytical way.

However, the hop rates of the misanthrope process realizing the TASEP with open boundaries, given in Table \ref{hoprate}, do not satisfy the condition which allows us to have the exact solution.
Alternatively, we modify the hop rates so as to have the exact steady-state probability and moreover to recover the misanthrope process with the hop rates in Table \ref{hoprate} (where we do not normalize $p$ to be 1 for explicitness).

\subsection{Condition for a factorized steady state}
We obtain the steady-state probability $P(\{n_l\})$ as a solution of the equation balancing probability currents from and into a given configuration $\{n_l\}$:
\begin{eqnarray}
\sum^L_{l=1} u(n_{l-1}+1,n_{l}-1)P(\{\ldots,n_{l-1}+1,n_{l}-1,\ldots\})\nonumber\\
=\sum^L_{l=1} u(n_{l},n_{l+1})P(\{\ldots,n_{l},n_{l+1},\ldots\}),
\label{mastereqn}
\end{eqnarray}
where we regard $u(m,-1)=0$, $n_0=n_L$ and $n_{L+1}=n_1$.
Assume that $P(\{n_l\})$ is given by a factorized form:
\begin{equation}
P(\{n_l\})=\frac1{Z_L}\prod^L_{l=1}f(n_l),
\label{FF}
\end{equation}
one finds a constraint on hop rates:
\begin{equation}
u(m,n)=u(m+1,n-1)\frac{u(1,m)u(n,0)}{u(m+1,0)u(1,n-1)}+u(m,0)-u(n,0),
\label{condition}
\end{equation}
and an exact solution of the weight function for single site:
\begin{equation}
f(n)=f(0)\Bigl(\frac{f(1)}{f(0)}\Bigr)^n\prod^{n}_{k=1}\frac{u(1,k-1)}{u(k,0)}\qquad (n\geq2).
\label{fn}
\end{equation}
Both $f(0)$ and $f(1)$ will remain undetermined, but however they do not change $P(\{n_l\})$ and consequently any expectation value.
These results were first given in \cite{Cocozza}.
See also \cite{EW} for further information and recent investigations.

\begin{figure}[htb]
\begin{center}
\includegraphics[scale=0.35]{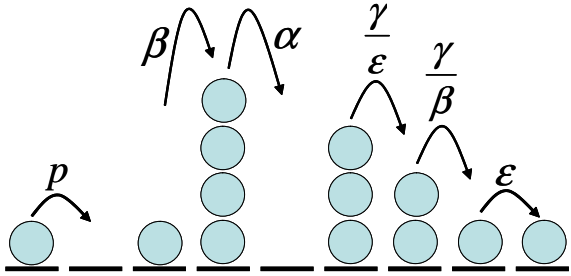}
\caption{The misanthrope process with the hop rates given in Table \ref{hoprate2}.
The hop rates given in Table \ref{hoprate}, which realize the TASEP with open boundaries, are recovered by letting $\gamma=\epsilon$ and taking the limit $\epsilon\to0$.}
\label{misanthrope}
\end{center}
\end{figure}

Now we modify the hop rates $u(m,n)$ in Table \ref{hoprate} to satisfy (\ref{condition}) so that we can have the exact steady state.
Since (\ref{fn}) includes only the hop rates $u(1,k\geq0)$ and $u(k\geq1,0)$, we firstly change the value of $u(1,1)$ from $0$ to $\epsilon$ then having the weight function:
\begin{equation}
f(n)=f(0)^{1-n}f(1)^{n}\epsilon\alpha^{1-n}\beta^{n-2}\qquad(n\geq2).
\label{f}
\label{fne}
\end{equation}
From the second row $u(1,*)$ and the second column $u(*,0)$, the constraint (\ref{condition}) determines the other hop rates $u(m\geq2,n\geq1)$ as follows:
\begin{enumerate}
\item $n=1$: the constraint reduces to $p-u(m,0)=u(1,m)-u(m,1)$.
(a)$m=1$: this case is trivial.
(b)$m\geq2$: $u(m\geq2,1)=\alpha+\beta-p$.
This implies $\alpha+\beta-p\geq0$.
\item $\displaystyle n=2$: the constraint reduces to $\alpha-u(m,0)=u(m+1,1)u(1,m)/\epsilon-u(m,2)$.
(a)$m=1$: $u(2,1)=\alpha+\beta-p$.
(b)$m\geq2$: $u(m\geq2,2)=(\alpha+\beta-p)\beta/\epsilon$.
\item $n\geq3$: the constraint reduces to $\alpha-u(m,0)=u(m+1,n-1)u(1,m)/\beta-u(m,n)$.
(a)$m=1$: $u(2,n\geq2)=(\alpha+\beta-p)\beta/\epsilon$.
(b)$m\geq2$: $u(m\geq3,n\geq2)=u(2,m+n-2)$.
\end{enumerate}
The above discussions are concluded in Table \ref{hoprate2}, where we let $\gamma=\beta(\alpha+\beta-p)$ for simple expression.
Note that we require $\gamma\geq0$.
In summary, we have the weight function (\ref{fne}) for the misanthrope process with the hop rates given in Table \ref{hoprate2}.
Figure \ref{misanthrope} illustrates the misanthrope process obtained here.
%%%%%%%%%%%
\begin{table}[htb]
\begin{center}
\begin{tabular}{c|ccccc}
$u(m,n)$ & $n=0$ & $n=1$ & $n=2$ & $n=3$ & $\cdots$ \\
\hline
$m=0$ & $0$ & $0$ & $0$ & $0$ & $\cdots$ \\
$m=1$ & $p$ & $\epsilon$ & $\beta$ & $\beta$ & $\cdots$ \\
$m=2$ & $\alpha$ & $\gamma/\beta$ & $\gamma/\epsilon$ & $\gamma/\epsilon$ & $\cdots$ \\
$m=3$ & $\alpha$ & $\gamma/\beta$ & $\gamma/\epsilon$ & $\gamma/\epsilon$ & $\cdots$ \\
$\vdots$ & $\vdots$ & $\vdots$ & $\vdots$ & $\vdots$ & $\ddots$
\end{tabular}
\caption{Hop rates for the modified misanthrope process which has a factorized steady-state probability, where $\gamma=\beta(\alpha+\beta-p)\geq0$.
}
\label{hoprate2}
\end{center}
\end{table}
%%%%%%%%%%%
%%%%%%%%%%%

Let $\gamma=\epsilon$ and then take the limit $\epsilon\to0$ in Table \ref{hoprate2}.
Thus we can recover the hop rates in Table \ref{hoprate}.
At the point of equating $\gamma$ with $\epsilon$, the misanthrope process allows for the exact solution, whereas it imposes an additional restriction for the independent parameters $p$, $\alpha$ and $\beta$:
\begin{equation}
\alpha+\beta=p+\epsilon.
\label{abpe}
\end{equation}
Consequently, in the limit $\epsilon\to0$ the misanthrope process with an exact steady state realizes the TASEP with the open boundary conditions under the restriction $\alpha+\beta=p$.

%%%%%%
\subsection{Current}
In the following we use the notations:
the configuration space $\Omega_{LN}=\{\omega=\{n_1,n_2,\ldots,n_L\}\mid \forall n_l\in\mathbf{Z}_{\geq0}, |\omega|=N\}$ where $|\omega|=\sum^L_{l=1}n_l$;
the number of sites $\sigma_k(\omega)$ carrying $k$ particles in a configuration $\omega$.
Note $\sum_{k\geq0}\sigma_k(\omega)=L$ and $\sum_{k\geq0}k\sigma(\omega)=N$.

Now we calculate the steady state probability $P(\omega)$.
From equation (\ref{fne}), we immediately have
\begin{eqnarray*}
\prod^L_{l=1}f(n_l)
&=&\prod_{k\geq0}f(k)^{\sigma_k(\omega)}\\
&=&f(0)^{L-N}f(1)^{N}\epsilon^{L-\sigma_0(\omega)-\sigma_1(\omega)}\\
&&\cdot\alpha^{L-N-\sigma_0(\omega)}\beta^{N-2L+2\sigma_0(\omega)+\sigma_1(\omega)}.
\end{eqnarray*}
Then, equation (\ref{FF}) gives $P(\omega)$
\begin{equation}
P(\omega)= \frac{f(0)^{L-N}f(1)^{N}\epsilon^{L-\sigma_0(\omega)-\sigma_1(\omega)}\alpha^{L-N-\sigma_0(\omega)}\beta^{N-2L+2\sigma_0(\omega)+\sigma_1(\omega)}}{Z_{LN}},
\label{P}
\end{equation}
with the normalization:
\begin{eqnarray}
Z_{LN}&=&\sum_{\omega\in\Omega_{LN}}\prod_{k\geq0}f(k)^{\sigma_k(\omega)} \nonumber\\
&=&f(0)^{L-N}f(1)^{N}\nonumber\\
&&\cdot \sum_{\omega\in\Omega_{LN}}\epsilon^{L-\sigma_0(\omega)-\sigma_1(\omega)}\alpha^{L-N-\sigma_0(\omega)}\beta^{N-2L+2\sigma_0(\omega)+\sigma_1(\omega)}.
\label{Z}
\end{eqnarray}

We rewrite the normalization (\ref{Z}) as
\begin{eqnarray}
Z_{LN}&=&f(0)^{L-N}f(1)^{N}F_{LN},\nonumber\\
F_{LN}&=&\sum_{\omega\in\Omega_{LN}}\epsilon^{L-\sigma_0(\omega)-\sigma_1(\omega)}\alpha^{L-N-\sigma_0(\omega)}\beta^{N-2L+2\sigma_0(\omega)+\sigma_1(\omega)}.
\label{ZF}
\end{eqnarray}
The current of particles on the lattice is defined by
%\begin{equation}
\[Q_{LN}=\sum_{\omega\in\Omega_{LN}}u(n_1,n_2)P(\omega).\]
%\end{equation}
Substitution of (\ref{P}) gives
\begin{eqnarray}
Q_{LN}&=&p \frac{F_{L-2,N-1}}{F_{LN}}+\epsilon \sum^N_{m=2}\Bigl(\frac{\beta}{\alpha}\Bigr)^{m-2} \frac{F_{L-2,N-m}}{F_{LN}}\nonumber\\
& & +\frac{\epsilon^2}\beta\sum^N_{m=1}\sum^N_{n=1}\Bigl(\frac{\beta}{\alpha}\Bigr)^{m+n-2} \frac{F_{L-2,N-m-n}}{F_{LN}},
\label{Q}
\end{eqnarray}
where we use (\ref{abpe}) to simplify the equation.

For further calculations, we shall count the number of the states $\omega$ such that $s_0=\sigma_0(\omega)$ and $s_1=\sigma_1(\omega)$.
Let $\Gamma_{LN}(s_0,s_1)$ denote the number of these states.
Then, $F_{LN}$ is rewritten as
\[F_{LN}=\sum_{s_0}\sum_{s_1} \Gamma_{LN}(s_0,s_1) \epsilon^{L-s_0-s_1}\alpha^{L-N-s_0}\beta^{N-2L+2s_0+s_1}.\]
Counting $\Gamma_{LN}(s_0,s_1)$ requires not difficult but careful discussions as follows.
\begin{enumerate}
\item $N>L$: in this case, it is obvious that $0\leq s_0\leq L-1$.
Denote by $\lambda_{LN}(s)$ the number of states such that $s_0=s$, and one finds
\[\lambda_{LN}(s)=\binom{L}{s} \binom{N-1}{L-s-1}.\]
Note that this formula is not available for the case of $N\leq0$ or $L<0$.
Consider that $\Gamma_{LN}(s_0,s_1)$ will be obtained from $\lambda_{L-s_0,N-(L-s_0)}(s_1)$:
\[\Gamma_{LN}(s_0,s_1)=\binom{L}{s_0}\lambda_{L-s_0,N-(L-s_0)}(s_1).\]
\item $N=L$:
if $s_0>0$ then this is the same case as $N>L$.
Hence we have
\[\Gamma_{NN}(s_0,s_1)=\binom{L}{s_0}\lambda_{N-s_0,s_0}(s_1).\]
If $s_0=0$, $\lambda_{N-s_0,s_0}(s_1)$ can not be defined.
In this case, one has a simple formula:
\[\Gamma_{NN}(0,s_1)=\delta_{s_1,N},\]
however, we can not include this in the previous formula.
\item $N<L$:
in this case, one sees $L-N\leq s_0\leq L-1$, and thus $0\leq N-L+s_0\leq N-1$.
If $s_0\ne L-N$, then
\[\Gamma_{LN}(s_0,s_1)=\binom{L}{s_0}\lambda_{L-s_0,N-L+s_0}(s_1).\]
If $s_0=L-N$, then
\[\Gamma_{LN}(L-N,s_1)=\binom{L}{L-N}\delta_{s_1,N}.\]
\end{enumerate}
In conclusion, we obtain
\begin{equation}
\Gamma_{LN}(s_0,s_1)=\left\{
\begin{array}{l}
\mbox{(i)} N>L:\ \binom{L-s_0}{s_1}\binom{N-L+s_0-1}{L-s_0-s_1-1}\\
\qquad (0\leq s_0\leq L-1, 0\leq s_1\leq L-1),\\
\mbox{(ii)} N\leq L:\\
\quad \mbox{(a)}s_0=L-N:\ \binom{L}{N}\delta_{s_1,N}\\
\qquad (0\leq s_1\leq N),\\
\quad \mbox{(b)}s_0>L-N:\ \binom{L-s_0}{s_1}\binom{N-L+s_0-1}{L-s_0-s_1-1}\\
\qquad (0\leq s_1\leq N),
\end{array}
\right.
\label{Gamma}
\end{equation}
where we define $\Gamma_{LN}(s_0,s_1)=0$ if $L-s_0< s_1\leq N$.

%%%%%%
\subsection{Current for the TASEP}
We continue to the next step to estimate the current (\ref{Q}) with respect to $\epsilon$.
In the following discussions, $\epsilon$ is assumed to be small enough.
Note that from (\ref{abpe}), $\alpha+\beta=p$ in the limit of $\epsilon\to0$.
The normalization $F_{LN}$, defined in (\ref{ZF}), is expanded to a series in $\epsilon$:
\begin{eqnarray}
F_{LN}&=&\sum^{L}_{l=0}\Gamma_{LN}(L-l,l)\alpha^{l-N}\beta^{N-l}\nonumber\\
&&+\epsilon\sum^{L-1}_{l=0}\Gamma_{LN}(L-l-1,l)\alpha^{l-N+1}\beta^{N-2-l}+O(\epsilon^2)\nonumber\\
&=&\left\{
\begin{array}{l}
\displaystyle
\epsilon L\alpha^{-N+1}\beta^{N-L-1}p^{L-1}+O(\epsilon^{2}) \qquad(N>L),\\
\displaystyle
\binom{L}{N}+\epsilon\sum^{N-2}_{l=0}L\binom{L-1}{l}\Bigl(\frac\beta\alpha\Bigr)^{N-l}+O(\epsilon^{2}) \qquad(N\leq L),
\end{array}
\right.
\label{F}
\end{eqnarray}
where we use (\ref{Gamma}) to see
\[
\Gamma_{LN}(L-l,l)=\left\{
\begin{array}{l}
\mbox{(i)} N>L:\ 0\qquad (0\leq l\leq L),\\
\mbox{(ii)} N\leq L:\\
\quad \mbox{(a)} l=N:\ \binom{L}{N},\\
\quad \mbox{(b)} l<N:\ 0,
\end{array}
\right.
\]
and
\[
\Gamma_{LN}(L-l-1,l)=\left\{
\begin{array}{l}
\mbox{(i)} N>L:\ L\binom{L-1}{l} \qquad (0\leq l\leq L-1),\\
\mbox{(ii)} N\leq L:\\
\quad \mbox{(a)} l=N-1:\ 0,\\
\quad \mbox{(b)} l<N-1:\ L\binom{L-1}{l}.
\end{array}
\right.
\]

In order to estimate the current $Q_{LN}$, given in (\ref{Q}), we expand $F_{L-2,N-m}/F_{LN}$ $(m=1,2,\ldots)$ to a series in $\epsilon$.
The series expansion of $F_{LN}$ in $\epsilon$ has a leading term of order $\epsilon^0$ if $N>L$, whereas of order $\epsilon^1$ if $N\leq L$.
Accordingly, we have some cases to consider:
\begin{enumerate}
\item[(A)] $N>L$: since obviously $N-L+2>2$, we calculate $F_{L-2,N-m}/F_{LN}$ in the following cases:
\[
\frac{F_{L-2,N-m}}{F_{LN}}
=\left\{
\begin{array}{l}
\displaystyle
\frac{L-2}L \alpha^m \beta^{-m+2}p^{-2}+O(\epsilon^{1})\\
\qquad (m<N-L+2),\\
\displaystyle
\frac1{\epsilon L} \binom{L-2}{N-m}\alpha^{N-1}\beta^{-N+L+1}p^{-L+1}+O(\epsilon^{0})\\
\qquad (m\geq N-L+2).
\end{array}
\right.
\]
\item[(B)] $N\leq L$: since $N-L-m+2< 2$ we have two cases $N-L-m+2=1$ and $N-m-(L-2)\leq0$; moreover, that $m>0$ and $N-L-m+2=1$ leads to $N=L$ and $m=1$:
\[\frac{F_{L-2,N-m}}{F_{LN}}
=\left\{
\begin{array}{l}
\mbox{(i)} N=L:\\
\quad\mbox{(a)} m=1:\
\displaystyle
O(\epsilon^{1})\\
\quad \mbox{(b)} m>1:\ 
\binom{L-2}{L-m}+O(\epsilon^1),\\
\mbox{(ii)} N<L:\
\displaystyle \frac{\binom{L-2}{N-m}}{\binom{L}{N}}+O(\epsilon^1).
\end{array}
\right.\]
\end{enumerate}
Substituting the formulas above to (\ref{Q}), we finally obtain the current $Q_{LN}$ expanded in $\epsilon$:
\begin{equation}
Q_{LN}=\left\{
\begin{array}{ll}
\displaystyle
\frac{L-1}L\frac{\alpha\beta}p+O(\epsilon^{1})\qquad&(N>L),\\
O(\epsilon^1)& (N=L),\\
\displaystyle
\frac{N(L-N)}{L(L-1)}p+O(\epsilon^1)&(N<L).\\
\end{array}
\right.
\label{Qe}
\end{equation}
We have distinct leading terms of $Q_{LN}$ according to the magnitude relationship between the number of particles $N$ and that of sites $L$.

In the case $N>L$, as the system reaches its steady state, more than one particle accumulate at a single site to build a {\em condensate}.
The condensate maintains itself by absorbing a particle from the left-hand site and providing one to the right-hand site.
%Thus, the remaining $L-1$ sites form the TASEP with open boundaries, where the condensate takes the role of the external to induce a particle current.
The condensate thereby takes the role of the external to induce a particle current in the lattice, and the remaining $L-1$ sites form the TASEP with open boundaries.
This may be referred to as a {\em fluid} comparing with a condensate.

The leading term of $Q_{LN}\ (N>L)$ in (\ref{Qe}) suggests the above scenario.
First, note that the particle current is given as $\alpha \beta/p$ for the TASEP normally defined in a lattice with left and right boundaries.
Note that in the random sequential updating, each site is chosen with probability $1/L$ to attempt an update of its state at every time unit.
In the present model, the TASEP comprises $L-1$ of the total $L$ sites in the lattice; hence the factor $(L-1)/L$ should multiply in (\ref{Qe}).

In the case $N=L$, we can easily imagine that all particles lie uniformly on the lattice, i.e., every site is occupied by just one particle, and there occurs very few hop of particles (at order $\epsilon$).
It may be called a {\em crystal}.

In the case $N<L$, the leading term in (\ref{Qe}) coincides with the current of the normal TASEP with the periodic boundary condition defined with $N$ particles on the lattice of $L$ sites \cite{Derrida98}.
Consequently, as far as $N$ is smaller than $L$ (even if $N=L-1$), we have no condensation or every condensate vanishes in the steady state and hence the TASEP with periodic boundary forms in the whole lattice.
The boundary rates $\alpha$ and $\beta$ no longer contribute to the dynamics in the steady state.

%%%%%%%%
\subsection{Condensation probability}
%As explained above, if there exist more particles than sites in the lattice, the surplus of particles spontaneously accumulates at a single site to form a condensate.
%In the steady state, the condensate behaves as a particle reservoir and thus the TASEP with open boundaries is realized over the remaining sites.
%Then, the current is entirely controlled by the boundary rates $\alpha$ and $\beta$, and these also determine the number of particles in the TASEP part of the lattice.
As mentioned above, a condensate behaves as a particle reservoir.
In the case $N>L$, at least one condensate arises and the TASEP with open boundaries is realized over the remaining sites in the lattice, where the current is entirely controlled by the boundary rates $\alpha$ and $\beta$.
Hence these rates determine the number of particles included in the condensate.
In this subsection, we calculate the probability $p_{LN}(n)$ that for any $L$ and $N$, $n$ particles accumulate at a single site in the steady state.

Any site will be chosen with the same probability $1/L$; one may take site 1 here.
The probability $p_{LN}(n)$ can be obtained from the steady-state probability (\ref{P}):
\begin{eqnarray*}
p_{LN}(n)&=&\sum_{\omega\in\Omega_{LN}(n_1=n)}P(\omega)
=f(n)\frac{Z_{L-1,N-n}}{Z_{LN}}\\
&=&\left\{\begin{array}{ll}
\displaystyle \frac{F_{L-1,N}}{F_{LN}} & (n=0),\\
\displaystyle \frac{F_{L-1,N-1}}{F_{LN}} & (n=1),\\
\displaystyle \epsilon \alpha ^{-n+1} \beta^{n-2} \frac{F_{L-1,N-n}}{F_{LN}} & (n\geq2),
\end{array}\right.
\end{eqnarray*}
where we use (\ref{f}) and (\ref{abpe}).
Then, substitution of (\ref{F}) yields
\begin{enumerate}
\item $N>L$:
\begin{eqnarray}
p_{LN}(0)=\frac{L-1}{L}\beta/p+O(\epsilon^1),\qquad p_{LN}(1)=\frac{L-1}{L}\alpha/p+O(\epsilon^1),\nonumber\\
p_{LN}(n\geq2)=\left\{
\begin{array}{l}
\mbox{(a)} 2\leq n\leq N-L:\ O(\epsilon^{1}),\\
\mbox{(b)} N-L+1\leq n\leq N:\\
\displaystyle \frac1L\binom{L-1}{N-n}\frac{ \alpha^{N-n}\beta^{n-(N-L+1)}}{p^{L-1}}+O(\epsilon^{1}).
\end{array}
\right.
\label{p(n)1}
\end{eqnarray}
\item $N=L$:
\[p_{LN}(0)=O(\epsilon^1),\quad p_{LN}(1)=1+O(\epsilon^1),\quad p_{LN}(n\geq2)=O(\epsilon^1).\]
\item $N<L$:
\begin{eqnarray}
p_{LN}(0)=\frac{L-N}L+O(\epsilon^1),\qquad p_{LN}(1)=\frac{N}L+O(\epsilon^1),\nonumber\\
p_{LN}(n\geq2)=O(\epsilon^1).\label{p(n)2}
\end{eqnarray}
\end{enumerate}

Ignore the terms of order $\epsilon$, and (\ref{p(n)1}) tells that there appears no condensate smaller than $N-L+1$.
However, this does not exactly conclude that no two condensates can coexist.
Nevertheless, if a condensate of more than $N-L+1$ particles grows at a single site, there is fewer than $L-1$ particles in the remaining $L-1$ sites.
From (\ref{p(n)2}), one consequently finds that there can be no other condensate in the lattice.

The leading term in (\ref{p(n)1}) corresponds to a binomial distribution, which implies that $p_{LN}(n)$ attains the maximum at $n=N-(L-1)\alpha/p$.
The condensate maintains about this height in the steady state, while the number of particles in the fluid part is $(L-1)\alpha/p$.

%%%%%%
\section{Conclusion}
%%%%%%
In this work, we propose a misanthrope process, defined on a ring, which realizes the totally asymmetric simple exclusion process (TASEP) with open boundaries.
In general, the misanthrope process has a high degree of freedom in making rules for particles hopping between sites, and is even exactly solvable if a constraint on the hop rates is satisfied.
We exploited these advantages to establish the present model and to confirm that it is successful.

Our idea is simple --- a condensate in a periodic lattice, if formed, can be a reservoir of particles providing and absorbing a particle.
In order to include the exclusion property of particles in the model, we let some hop rates be zero.
(This may be a point to discuss.)

It is striking that the TASEP shows boundary-induced phase transitions, i.e., the boundary rates with which a particle gets into and out of the lattice;
some change of the boundary rates leads to drastic changes of the particle current.
In the present model, the same phenomena take place according to the number of particles: if it exceeds the number of sites, a condensate remains at a single site and then behaves as a particle reservoir.
If it is smaller than the number of sites, a homogeneous current of particles circulates in the lattice.
If these numbers are equal, each particle is trapped in a site by the exclusive interaction with the other ones and the whole system results in gridlock.
It is interesting that each of these outstanding phenomena appears in one system being selected by the density of particles.

Exact solution of the misanthrope process enables us, under the solvability condition on the hop rates, to have the particle current and the condensation probability in analytical form.
Thus we make sure that, in particular, condensation occurs at a single site in the case that the particle density is more than one.
Actually, it will require much computation time to find if some condensates existing in early steps merge into one at last, as the system size becomes large.

The solvability condition of the misanthrope process leads to a factorized form of the steady-state probability, when in fact correlations between two adjacent sites vanish and {\em non-commutative} matrix products for the steady state accordingly reduce to a commutative product (i.e., scalar) of the weight functions each for a single site.
%Conversely, we already have an exact solution of the misanthrope process realizing the TASEP with open boundaries.
%Can this fact give a further possibility to the misanthrope process?
Conversely, can the matrix-product ansatz suggest another possibility of the misanthrope process?
It will be a future work.

%%%%%%%%%%%%%
\begin{acknowledgements}
This work was supported in part by JSPS KAKENHI Grant Number 26610033.
%Grant-in-Aid for Challenging Exploratory Research Number 26610033.
\end{acknowledgements}


%\begin{thebibliography}{}
% etc
%\end{thebibliography}

\end{document}